\documentclass[DIV9,a4paper,final]{scrartcl}

\usepackage{microtype}
\usepackage[utf8]{inputenc}
\usepackage{amsmath}
\usepackage{amssymb}
\usepackage{stmaryrd}
\usepackage{color}
\usepackage{xspace}

\SetSymbolFont{stmry}{bold}{U}{stmry}{m}{n}

\newcommand{\set}[2]{\left\{#1\mathrel{\left|\vphantom{#1}\vphantom{#2}\right.}#2\right\}}
\newcommand{\oneset}[1]{\left\{\mathinner{#1}\right\}}

\newcommand{\os}[1]{\oneset{#1}}

\newcommand{\Oh}{\mathcal{O}}

\newcommand{\abs}[1]{\left|\mathinner{#1}\right|}

\newcommand{\N}{\mathbb{N}}
\newcommand{\Z}{\mathbb{Z}}

\renewcommand{\phi}{\varphi}

\newcommand{\gam}{\gamma}

\newcommand{\lam}{\lambda}

\newcommand{\cA}{\mathcal{A}}

\newcommand{\cL}{\mathcal{L}}
\newcommand{\cR}{\mathcal{R}}
\newcommand{\cH}{\mathcal{H}}

\newcommand{\cD}{\mathcal{D}}

\newcommand{\cG}{\mathcal{G}}

\newcommand{\cJ}{\mathcal{J}}

\newcommand{\cP}{\mathcal{P}}

\newcommand{\bfAb}{\mathbf{Ab}}

\newcommand{\bfI}{\mathbf{1}}

\newcommand{\bfH}{\mathbf{H}}

\newcommand{\bfMon}{\mathbf{Mon}}

\newcommand{\bfG}{\mathbf{G}}

\newcommand{\bfV}{\mathbf{V}}

\newcommand{\ov}[1]{\overline{#1}}
\newcommand{\oi}[1]{{#1}^{-1}}

\newcommand\subsem{subsemigroup\xspace}

\newcommand{\ie}{i.e.,\xspace}

\newcommand{\IFF}{if and only if\xspace}
\newcommand{\homo}{homomorphism\xspace}

\newcommand{\var}{variety\xspace}

\newcommand{\SF}{\mathrm{SF}}
\newcommand{\AP}{\mathbf{Ap}}

\newcommand{\schuetz}{Schützenberger\xspace}

\newcommand{\sse}{\subseteq}
\newcommand{\ssneq}{\varsubsetneq}
\newcommand{\es}{\emptyset}
\newcommand{\sm}{\setminus}

\newcommand{\SD}{\ensuremath{\mathrm{SD}}}

\newcommand{\SDH}[1]{\ensuremath{\mathrm{SD}_{#1}}}

\newcommand{\LocRees}[1]{\ensuremath{\mathrm{LocRees(#1)}}}
\newcommand{\Rees}[1]{\ensuremath{\mathrm{Rees(#1)}}}

\definecolor{lfgreen}{rgb}{0.2,0.8,0.4}
\newenvironment{lf}{\noindent\color{lfgreen}LF}{} 

\newenvironment{vd}{\noindent\color{blue}VD}{}

\newcommand{\malcev}{\mathop{\text{\textcircled{\scriptsize M}}}}

\title{Church-Rosser Systems, Codes with Bounded Synchronization Delay and Local Rees Extensions}

\author{Volker Diekert \and Lukas Fleischer\thanks{Supported by the German Research Foundation (DFG) under grant DI 435/6-1.}}

\date{FMI, University of Stuttgart\\
  Universitätsstraße 38, 70569 Stuttgart, Germany\\
  \normalsize\texttt{$\{$diekert,fleischer$\}$@fmi.uni-stuttgart.de}}

\begin{document}

\maketitle

\begin{center}
  \emph{In memoriam: Zolt\`an \'Esik (1951 -- 2016)}
\end{center}

\begin{abstract}
What is the common link, if there is any, between Church-Rosser systems, prefix codes with bounded synchronization delay, and local Rees extensions? The first obvious answer is that each of these notions relates to topics of interest for WORDS: Church-Rosser systems are certain rewriting systems over words, codes are given by sets of words which form a basis of a free submonoid in the free monoid of all words (over a given alphabet) and local Rees extensions provide structural insight into regular languages over words. So, it seems to be a legitimate title for an extended abstract presented at the conference WORDS 2017.
However, this work is more ambitious, it outlines some less obvious but much more interesting link between these topics. This link is based on a structure theory of finite monoids with varieties of groups and the concept of \emph{local divisors} playing a prominent role. Parts of this work appeared in a similar form in conference proceedings \cite{DiekertKRW12icalp,DiekertWalter16} where proofs and further material can be found. 
\end{abstract}

\section*{Introduction}
\label{sec:intro}

\emph{Ceci n'est pas une introduction.\footnote{Following  ``La trahison des images'' by  Ren\'e Magritte.}} The present paper does not claim to provide any new results. Its purpose is to give an overview on a theory developed over the past twenty years, having its origins in a construction derived from the \emph{Habilitationsschrift} of Thomas Wilke \cite{wil98} which the first author was refereeing in 1997. Inspired by this construction (which also appears in \cite{wil99stacs}), he distilled the concept of a {local divisor} of a finite monoid without knowing that this concept existed long before in commutative algebra \cite{Mey72} (denoted by Kurt Meyberg as \emph{local algebra}) and without giving any special name to it. The term \emph{local divisor} was coined 2012 in \cite{DiekertKS12fi} only.\footnote{The pointer to \cite{Mey72} is due to Benjamin Steinberg and that a ``local divisor'' is a monoid divisor in the usual sense was observed by Daniel Kirsten, first. Thanks!}

Originally, the concept was solely used as a tool to simplify existing proofs. Still, this was particularly helpful in \cite{dg06IC} which  introduced
this proof technique to the semigroup community. However, over the last decade, it gave rise to new results. Amazingly, it was powerful enough to solve long-standing open problems.  It is hard to formally pinpoint where the  power of method comes from or why, on the other hand, it has clear limitations. Let us conclude with an \emph{\'etale} statement: There are not enough local submonoids, so the role of local submonoids transfers to local divisors and there are plenty of them. This seems to be useful.

\section*{Preliminaries}
\label{sec:prel}

Throughout the paper $A$ denotes a finite alphabet and $M$, $N$ denote monoids. If not stated otherwise, $M$ and $N$ will be finite.
A \emph{divisor} of a monoid $M$ is a monoid $N$ which is a homomorphic image of a \subsem of $M$.
A \emph{variety} of finite monoids is a nonempty family of finite monoids $\bfV$ which is closed under taking divisors and finite direct products.
A variety of finite groups is a variety of finite monoids where each of the monoids is a group.

The largest group variety is $\bfG$, the \var of all finite groups.
If $\bfH$ is a variety of finite groups, $\ov \bfH$ denotes the class of finite monoids where all subgroups are members of $\bfH$. It turns out that for every group variety $\bfH$, the class $\ov \bfH$ is a variety, see~\cite{eil76}. Actually, it is the greatest variety of finite monoids such that $\ov \bfH \cap \bfG = \bfH$.
 Clearly, $\ov \bfG$ is the class of all finite monoids which we denote by  $\bfMon$.
 The most prominent subclass is $\ov \bfI$, the variety of aperiodic monoids $\AP$. Here, $\bfI$ denotes the smallest group variety, containing the trivial group $\os 1$ only. 

Given a \var $\bfV$, we denote by $\bfV(A^*)$ the set of languages 
$L\sse A^*$ such that $L=\oi\phi(\phi(L))$ for some \homo $\phi:A^*\to M$ where $M\in \bfV$. From formal language theory, we know that 
$\bfMon(A^*)$ is the set of all regular languages in $A^*$. 

\section{Church-Rosser Thue Systems}\label{sec:crts}

A \emph{semi-Thue system} is a set of rewriting rules $S\sse A^*\times A^*$ over some alphabet $A$. (For simplicity, throughout this paper, semi-Thue systems are assumed to be finite.) A system $S$ defines a finitely presented quotient monoid $$A^*/S= A^*/\set{\ell=r}{(\ell,r)\in S},$$
and the system is called \emph{Church-Rosser} (with respect to the length function) if $S$ is confluent and  length-reducing. The interest in Church-Rosser systems stems from the fact that we can compute irreducible normal forms in linear time (as the system is finite and length-reducing) and that the irreducible 
normal forms of two words $u,v$ are identical \IFF $u$ and $v$ represent the same word in $A^*/S$ (as the system is confluent). Thus, if a monoid $M$ has a presentation as 
$M=A^*/S$, then the word problem of $M$ is solvable in linear time. 
The notion of a \emph{Church-Rosser language} is an offspring of that observation and  appeared 
first in Narendran's PhD thesis~\cite{Narendran84phd}, followed by a systematic study of that concept in~\cite{McNaughtonNO88}. 
As a result, \cite{McNaughtonNO88} defines a language class strictly larger than the 
the class of deterministic context-free languages for which the word problem is 
solvable in linear time. The authors of this work also define a restricted class which is incomparable with the class of (deterministic) context-free languages. 

A language $L\sse A^*$ is called \emph{Church-Rosser
congruential}, if there exists a finite, confluent, and length-reducing
semi-Thue system $S\sse A^* \times A^*$ such that $L$ is a finite union of congruence classes modulo~$S$. If, in addition, the index of $S$ is finite (\ie the monoid $A^*/S$ of all congruence classes is finite) then $L$ is called  \emph{strongly}  Church-Rosser congruential. Strongly Church-Rosser congruential languages are necessarily regular. It was conjectured (but open for more than 25 years until 2012) that all regular languages are (strongly) Church-Rosser congruential.
Some partial results were known before 2012 but commutativity in the syntactic monoid seemed to be a major obstacle. For example, it is easy to verify the conjecture provided the syntactic monoid is a finite non-Abelian simple group like $\cA_5$. On the other hand, it is surprisingly hard to prove the result for the Klein group $\Z/2\Z\times \Z/2\Z$.
Nevertheless, \cite{DiekertKRW15jacm} proved a
stronger result. Given a regular language $L\sse A^*$ and any weight function $\gam:A\to \N\sm \os{0}$, there exists a finite confluent and weight-reducing semi-Thue system $S$ such the quotient monoid is $A^*/S$ is finite and such that $L$ is a (necessarily finite) union of congruence classes.
This result is indeed stronger because the mapping $w \mapsto \abs{w}$ is just one particular weight function.

\section{Star-Freeness and Bounded Synchronization Delay}\label{sec:sfbs}

The class of \emph{star-free languages} over some alphabet $A$, denoted by $\SF(A^*)$, is the least class of languages which contains all finite languages over $A$ and which is closed under both Boolean operations (finite union and complementation) and concatenation. As the name suggests, we do not allow the Kleene star. 
Nevertheless, $B^*$ is star-free for all $B\sse A$. 
A fundamental result of Sch\"utzenberger characterizes the class of star-free languages by aperiodic monoids~\cite{sch65sf}. That is, a regular language belongs to $\SF(A^*)$ \IFF all subgroups in its syntactic monoid are trivial.
By slight abuse of notation, one usually abbreviates this result by $\SF = \AP$ as a short version of $\SF(A^*) = \AP(A^*)$.
Sch\"utzenberger found another, but less prominent characterization of $\SF$: the star-free languages are exactly the class of languages which can be defined inductively by finite languages and closure under finite union, concatenation, and the Kleene star restricted to prefix codes of bounded synchronization delay~\cite{Schutzenberger1975d}. This result is abbreviated by $\SD = \AP$. 

A language $K \sse A^+$ is called \emph{prefix code} if it is \emph{prefix-free}, \ie $u \in K$ and $uv \in K$ implies $u=uv$. A prefix-free language $K$ is a code since every word $u \in K^*$ admits a unique factorization $u = u_1 \cdots u_k$ with $k \geq 0$ and $u_i \in K$. 
A prefix code $K$ has \emph{bounded synchronization delay} if for some
$d \in \N$ and for all $u,v, w \in A^*$ with $uvw\in K^*$ and $v \in K^d$, we have $uv \in K^*$.
Note that the condition implies that for all $uvw\in K^*$
with $v \in K^d$, we have $w\in K^*$, too.
The idea is as follows: assume that a transmission of a code message is interrupted and we receive a fragment of the form $u''vw$ where $v \in K^d$ and $w\in K^*$. Then, we know that the original message was of the form 
 $u'u''vw$ with $u'u''v\in K^*$ and $w\in K^*$. Hence, we can decode $w$ as part of the original message. With a delay of $d$ code words the decoding can be synchronized. For $B\sse A$ and $c\in A\sm B$, the star-free language $B^*c$ is a prefix code of delay $1$ and $(B^*c)^+= (B\cup \os c)^*c$  
is star-free. The block code $A^2$ is finite, but not of bounded synchronization delay. Moreover, $(A^2)^*$ is not star-free as its syntactic monoid is the cyclic group of order two. 

Sch\"utzenberger's result $\AP \sse \SD$ is actually stronger than the well-known $\SF = \AP$ because proving the inclusions $\SD \sse \SF \sse \AP$ is relatively easy, see~\cite[Chapter VIII]{pp04}, so  $\SF = \AP$ follows from $\AP \sse \SD$.
A simple proof for $\AP = \SD$ including an extension to infinite words (which was not known before) was obtained much later in \cite{DiekertKufleitner14tocs}. It could be achieved thanks to the same algebraic decomposition into submonoids and local divisors.

\section{Local Divisors}\label{sec:locdiv}

In this section $e\in M$ denotes an idempotent, that is $e^2=e$. For such an idempotent, the set $M_e = eMe$ forms a monoid with $e$ as the identity element.
It is called the \emph{local monoid} at $e$. A \emph{local divisor} generalizes this concept by considering any element $c\in M$ 
and the set $M_c= cM\cap Mc$. Note that $eMe=eM\cap Me$, so local monoids are indeed a special case of local divisors. The next step is to define a multiplication $\circ$
on $cM\cap Mc$ by letting
 $$xc \circ cy = xcy$$
for all $x, y \in M$.
A straightforward calculation shows that the structure 
$(M_c,\circ,c)$ defines a monoid with this operation where the
neutral element of $M_c$ is $c$. This works for every $c\in M$. If $c$ is a unit, then $M_c$ is isomorphic to $M$. If $c$ is idempotent, then $M_c$ is the local monoid at the idempotent $c$. However, in general $M_c$ does not appear as a subsemigroup in $M$. 

The important fact is that $M_c$ is always a divisor of $M$.
Indeed, the mapping $\lam_c : \set{x\in M}{cx \in Mc} \to M_c$ given by $\lam_c(x) = cx$ is a surjective \homo. Moreover, 
if $c$ is not a unit, then $1 \notin cM \cap Mc$, hence 
$M_c\ssneq M$. This makes the construction suitable for induction. 

\section{Rees Extensions}\label{sec:Rees}

Let $N, L$ be monoids and $\rho: N \to L$ be any mapping. 
The \emph{Rees extension} over  $N,L,\rho$ is a classical construction
for monoids \cite{pin86,rs09qtheory}, frequently described in terms of matrices. It was used in the synthesis theory of Rhodes and Allen \cite{RhodesA73} which says that we can represent every finite monoid as a divisor of iterated Rees extensions, starting with groups. The ``advantage'' is that starting with a variety of groups $\bfH$, the construction produces monoids in $\ov \bfH$, only. This is not true for taking wreath products which are used in Krohn-Rhodes theory. For example, the symmetric group over three elements is not nilpotent, but appears as a subgroup of the wreath product of $\Z/3\Z$ and $\Z/2\Z$. Our definition of a Rees extension is similar, but not the same as the classical one. It avoids matrices and it is exactly as in \cite{DiekertKW12tcs}.  
The carrier set is
$$\Rees{N,L,\rho} = N \cup \left(N\times L \times N\right).$$ 
Let $n_1,n_1',n_2,n_2'\in N$ and $m,m'\in L$. Then the multiplication $\cdot$ on $\Rees{N,L,\rho}$ is given by 
\begin{align*}
n\cdot n' &= nn',\\
n\cdot (n_1,m,n_2) \cdot n' &= (nn_1, m , n_2n') ,\\
(n_1,m,n_2)\cdot (n_1',m',n_2') &= (n_1,m\rho(n_2n_1')m',n_2').
\end{align*}
For  a \var $\bfV$ of finite monoids let $\Rees{\bfV}$ be the least variety which contains $\bfV$ and which is closed under Rees extensions $\Rees{N,L,\rho}$. 
Almeida and Kl{\'i}ma called a \var $\bfV$ \emph{bullet-idempotent} if $\bfV = \Rees{\bfV}$, see~\cite{AlmeidaK16}. They showed 
$\Rees{\bfV} \sse {\ov \bfH}$ where $\bfH= \bfV\cap \bfG$ and asked whether all bullet-idempotent varieties are of that form. 
The answer is ``yes'' \cite{DiekertWalter16} and can be proved by showing the stronger result that so-called \emph{local Rees extensions} suffice to capture all of ${\ov \bfH}$. To define these objects, consider a finite monoid $M$ (which is not a group), an element 
$c\in M$, and a smaller submonoid $N$ of $M$ such that $N$ and $c$ generate $M$. Then, let $M_c$ be the local divisor at $c$ and let $\rho_c$ be  the mapping $\rho_c:N\to M_c$ with $\rho_c(x) = cxc$.
The \emph{local Rees extension} $\LocRees{N,M_c}$ is defined as the Rees extension $\Rees{N,M_c,\rho_c}$. Thus, a local Rees extension is a special case of a Rees extension. Still the result in \cite{DiekertWalter16}  shows that ${\ov \bfH}=\LocRees{\bfH}$. Here, $\LocRees{\bfH}$ denotes the least variety which contains the group variety $\bfH$ and which is closed under local Rees extensions. Since $\Rees{\bfV} \sse {\ov \bfH}$ for  $\bfH= \bfV\cap \bfG$ we obtain 
$$\Rees{\bfV} \sse {\ov \bfH} = \LocRees{\bfH} \sse \Rees{\bfV}.$$
Hence, all varieties appearing in the line above coincide.

\section{The Local Divisor Technique and Green's Lemma}\label{sec:ldt}

For a survey on the local divisor technique we refer to \cite{DiekertK2015tcs}. 
In general, there are more local divisors than local monoids, so having information about the structure in all local divisors tells us more about the structure of $M$ than just looking at the local monoids. Before we 
continue let us revisit Green's Lemma as sort of a ``commercial break'' for the local divisor technique in semigroup theory. 

The following section is based on~\cite{CostaS14forum,edam16} and
closely follows the presentation in~\cite[Cor.~7.45]{edam16} where full proofs are given.
Green's relations are classical. 
There are three basic equivalence relations $\cL$, $\cR$, and $\cJ$ which relate elements in a monoid $M$ generating the same left- (resp.~right-, resp.~two-sided-) ideal.
\begin{alignat*}{20}
  x &\cL y & \;\iff\; Mx &\,=\, & My, & \quad
  x &\cR y & \;\iff\; xM &\,=\,& yM, & \quad
  x &\cJ y & \;\iff\; MxM &\,=\,& MyM.
\end{alignat*}
The other two relations are defined by $\cH = \cL \cap \cR$ and $\cD = \cL \circ \cR$.
In particular,  
 $$x \cD y \iff \exists z: x \cL z \wedge z \cR y.$$ 
A standard exercise shows that $\cJ= \cD$ for finite monoids. (For infinite monoids this false, in general.) As $\cJ$ is symmetric, $\cJ= \cD$ implies 
$\cD = \cL \circ \cR= \cR \circ \cL$. The latter assertion is independent of that: $\cL \circ \cR= \cR \circ \cL$ holds  in infinite monoids, too. Therefore, all relations above are equivalence relations. If $\cG$ is any of them and $s\in M$, then
we write $\cG(s)=\set{t\in M}{s \cG t}$ for the equivalence class of $s$.

In the following, we assume that $M$ is finite. 
If $G$ is a subgroup of $M$ with neutral element $e$, then $G$ is a subgroup in $\cH(e)$; and $\cH(e)$ itself is a group. 
Now, Green's Lemma says that the groups $\cH(e)$ and $\cH(f)$ are isomorphic if $e$ and $f$ are idempotents belonging to the same $\cD$-class. The classical proof uses $\cD = \cL \circ \cR$. Hence, 
$e\cR z \cL f$ for some $z\in M$. Then one shows that 
the  right multiplication ${\cdot} v$, 
mapping $x$ to $xv$, induces a bijection $Me \to Mz$, $x\mapsto xv$.
By symmetry, we obtain a bijection between $\cH(e)$ and $\cH(f)$ which turns out to be an isomorphism of groups.

The proof is somewhat ``mysterious'' because the isomorphism passes through  $\cH(z)$ which is not subgroup of $M$, in general. Using local divisors however, the proof becomes fully transparent and reveals a more general fact. For this, consider any two $\cR$-equivalent (or symmetrically $\cL$-equivalent) elements $s$ and $t$. 
Whether or not $s$ or $t$ are idempotent, we can can define the local divisors $M_s$ and $M_t$. For  $s\cR t$ we can write $t= sv$ and now, the 
right multiplication $\cdot v$ defines an isomorphism $M_s \to M_t$. Moreover, as a set, $\cH(s)$ is the group of units in  $M_s$. In the case that $s=e$ is an idempotent $M_s=M_e$ is a local monoid and $\cH(s)= \cH(e)$ is a subgroup of $M$. Thus, as in the scenario of $e\cR z \cL f$ with idempotents $e$ and $f$ we see that three groups are isomorphic:
$\cH(e)$, $\cH(z)$ as the group of units in $(zM\cap Mz, \circ, z)$, and 
$\cH(f)$. There is no mystery in Green's Lemma if we view it from a more general perspective. 

\section{The Common Theme: Local Divisor Proofs}\label{sec:ct}

Let us now discuss the common theme in Church-Rosser systems, bounded synchronization delay, and Rees extensions. {}From an abstract viewpoint these deal with properties $\cP$ which can be defined for regular languages. Assume we know that a property $\cP$ of regular languages is true for all languages 
where the syntactic monoid belongs to some variety of groups $\bfH$. 
Then $\cP$ holds for all languages where the syntactic monoid belongs to $\ov \bfH$ \IFF and we can show the following implication for local Rees extensions $\LocRees{N,M_c}$:
\begin{align}\label{eq:rloc}
\cP(N) \wedge \cP(M_c) \implies \cP(\LocRees{N,M_c}).
\end{align}
Actually, it is enough show an implication without  mentioning
$\LocRees{N,M_c}$:
\begin{align}\label{eq:loc}
\cP(N) \wedge \cP(M_c) \implies \cP(M).
\end{align}
The reason that we mention the ``complicated'' implication (\ref{eq:rloc})
 is that the power of the method lies in the underlying algebraic connection between $N$, $M_c$ and $M$ which is best reflected by the local Rees extension. For simplicity of notation we just focus on the equivalent condition (\ref{eq:loc}). This
 implication is particularly appealing for aperiodic monoids. Indeed, any nontrivial property which is closed under taking submonoids must also hold for the trivial group $\os 1$. 
So, the base for the induction is trivial for the variety $\bfI$. In order 
to prove that $\cP$ holds for all aperiodic languages, one only needs to show~(\ref{eq:loc}). 
Sometimes this is very easy. Remember $\SF=\AP$, the probably most cited result of 
Sch\"utzenberger. The inclusion $\SF\sse\AP$ is rather straightforward and the assertion $\bfI(A^*)\sse\SF(A^*)$ is trivial since $\es$ and its complement $A^*$ are star-free. Now proving, (\ref{eq:loc}) is possible within less than a page, see \cite{Kufleitner14DCFS}. Almost the same holds for the less famous but more general result $\AP = \SD$, see~\cite{DiekertKufleitner14tocs}.

What about Krohn-Rhodes theory? It goes beyond $\AP$, but the group case is built-in!
The theory says that every monoid can be constructed by iterated wreath products, starting from finite simple groups and the so-called \emph{reset monoid} $U_2$.  According to \cite[page 241]{rs09qtheory} the monoid $U_2$ is ``essentially junk'' whereas the ``groups are gems''. Showing (\ref{eq:loc}) for the Krohn-Rhodes property 
was done in \cite{DiekertKS12fi} and led to a surprisingly easy proof of the Krohn-Rhodes decomposition theorem. 

Returning to prefix codes of bounded synchronization delay, it is worth mentioning that Sch\"utzenberger did not stop this line of research by showing that $\AP = \SD$. In~\cite{Schutzenberger1974b} he was able to prove an analogue of $\AP = \SD$ for languages where syntactic monoids have Abelian subgroups, only. For several years, no such characterization was known beyond $\ov \bfAb$. 

\subsection{\schuetz{}'s {$\SD$} Classes}\label{sec:sd}

Let $\bfH$ be a variety of finite groups. 
Consider a prefix code $K$ with bounded synchronization delay which can be written as a disjoint union $K=\bigcup\set{K_g}{g\in G}$ where $G\in \bfH$ and each $K_g$ is regular in $A^*$. The \emph{$\bfH$-controlled star} (more precisely, the 
\emph{$G$-controlled star}) associates with such a disjoint union the following language: 
$$\set{u_{g_1}\cdots u_{g_k}\in K^*}{u_{g_i}\in K_{g_i}\wedge g_1 \cdots {g_k}= 1 \in G}.$$ 
Another view of the $G$-controlled star of $K$ is the following: Let $\gamma_K : K \to G$ be a mapping such that $K_g = \gamma_K^{-1}(g)$ and let $\gamma: K^* \to G$ denote the canonical extension of $\gamma_K$ to a homomorphism from the free submonoid $K^* \sse A^*$ to $G$, then the 
$G$-controlled star of $K$ is exactly the set $\gamma^{-1}(1)$. 
Let $\mathcal C$ be any class of languages. We say that $\mathcal C$ is closed under $\bfH$-controlled star if for all $K$ and for every group $G \in \bfH$, the following closure property holds: if $K=\bigcup\set{K_g}{g\in G}$ is a prefix code with bounded synchronization delay such that $K_g \in \mathcal C$ for all $g \in G$, then the $G$-controlled star $\gamma^{-1}(1)$ is in $\mathcal C$ as well.
By $\SDH{\bfH}(A^*)$ we denote the smallest class of regular languages containing all finite subsets of $A^*$ and being closed under finite union, concatenation, and $\bfH$-controlled star.

Note that the definition of $\SDH{\bfH}(A^*)$ does not use any complementation.
Using different notation, Sch\"utzenberger showed that $\SDH{\bfH}(A^*) \sse \ov {\bfH}(A^*)$ in~\cite{Schutzenberger1974b}, but he proved the converse inclusion only for $\bfH \subseteq \bfAb$. The main result in~\cite{DiekertWalter16} states that
$\SDH{\bfH}(A^*) = \ov {\bfH}(A^*)$ for all $\bfH$. In retrospective, it is hard to say why Sch\"utzenberger did not prove this general result.
Perhaps he was not interested in that, but we believe that this is unlikely because he proved half of it. More likely, he tried to use the Krohn-Rhodes decomposition as in \cite{Schutzenberger1974b} which involves wreath products and they may take you outside 
$\ov \bfH$. Perhaps, Krohn-Rhodes theory was simply the wrong tool for this result. Local Rees extensions, on the other hand, are perfectly suitable for this kind of applications.

\subsection{Church-Rosser Thue Systems Revisited}\label{sec:crrev}

In the following $M$, denotes a finite monoid.
The results in Section~\ref{sec:crts} have their origins in formal language theory and led to the notion of Church-Rosser congruential languages. As mentioned before, for more than 25 years it was open whether or not all regular languages are  Church-Rosser congruential. A positive answer was given in~\cite{DiekertKRW15jacm}, and the corresponding  theorem has a purely algebraic formulation. It says that for each \homo $\phi$ from $A^*$ to $M$ factorizes through $A^*/S$ where $S$ is a finite confluent and length-reducing semi-Thue system of finite index. Thus, $\phi(\ell)=\phi(r)$ and $\abs \ell > \abs r$ for all $(\ell,r)\in S$. Moreover, $A^*/S$ is a finite monoid. 

For the inductive argument, one crucial idea is to consider weight functions $\gam:A\to \N\sm \os 0$.  The statement then becomes ``for every weight function and every every homomorphism $\phi:A^* \to M$ there exists  a finite confluent  semi-Thue system $S$ of finite index
such that $\phi(\ell)=\phi(r)$ and $\gam(\ell) > \gam(r)$ for all $(\ell,r)\in S$''. Instead of weight-reducing systems we can also define the notions of Parikh-reducing and subword-reducing systems. For a letter $a$ and 
a word $w\in A^*$ we let $\abs{w}_a$ be the number of $a$'s which occur in $w$. This defines a canonical \homo $\pi:A^*\to \N^A$ by $\pi(w) =( a \mapsto \abs{w}_a)$. The vector $\pi(w)$ is usually called the \emph{Parikh-image} of $w$. We say that $S$ is \emph{Parikh-reducing}
if $(\ell,r)\in S$ implies 
$\abs{\ell}_a\geq \abs{r}_a$ for all $a\in A$ and $\abs{\ell}_a> \abs{r}_a$
for at least one $a\in A$. Clearly, a \emph{Parikh-reducing} system is 
weight-reducing for every weight function. In the following, when using the term ``subword'' we mean ``scattered subword''. More precisely, a word $u$ is called a \emph{subword} of $w$ if there exists a factorization $u=a_1\cdots a_k$ such that $w \in A^* a_1 A^* \cdots a_k A^*$.  We say that $S$ is \emph{subword-reducing}
if $(\ell,r)\in S$ implies $\ell \neq r$ and that $r$ is a subword of $\ell$. Clearly, a subword-reducing system is Parikh-reducing.
The induction scheme (\ref{eq:loc}) introduced in the beginning of Section~\ref{sec:ct} works for all variants, but the group case is quite different. The trivial group leads to the subword-reducing system $\set{(a,1)}{a\in A}$. Consequently, the result in~\cite{DiekertKW12tcs} speaks about subword-reducing systems and this is the strongest result. The PhD thesis of Tobias Walter \cite{Walter16diss} shows that for 
all \homo{s} to 
Abelian groups there exists a 
Parikh-reducing Church-Rosser system as desired, thereby allowing him to construct such systems for all languages in $\overline{\bfAb}$.
Additionally, he proves that for all regular languages $L$ over a two letter alphabet there exists a Parikh-reducing Church-Rosser system $S$ of finite index such that $L$ is recognized by $A^*/S$. This shows that the existence of Parikh-reducing presentations is not limited to the \var $\overline{\bfAb}$.

\section{Conclusion and Open Problems}\label{sec:cop}

This extended abstract deals with the recurring theme of proving results for varieties of finite monoids and their associated language classes.
The most prominent example is the \var $\AP$ of aperiodic monoids, but our methods go beyond.
We have seen deep connections between apparently quite different objects where the technique allows to transfer results from a group \var $\bfH$ to its closure $\ov \bfH$.

Let us conclude with some open problems, starting with the new perspective on
Church-Rosser systems given in the previous subsection.

For subword-reducing and Parikh-reducing Church-Rosser systems, only partial
results are known. To date, it is still open whether subword-reducing
(resp.\ Parikh-reducing) Church-Rosser systems exist for every regular language.
It is tempting to believe that Parikh-reducing systems exist for all regular
languages, but we refrain from any conjecture in this case.

The notion of local Rees extensions gives rise to various interesting combinatorial problems concerning the complexity of Rees decompositions.
For a finite monoid $M$, a \emph{Rees decomposition tree} of $M$ is a rooted node-labeled tree such that the following conditions are satisfied.
\begin{itemize}
  \item The root has label $M$.
  \item Every inner node with label $M'$ has two children labeled by $N, M_c'$  such that $M'$ is a divisor of the local Rees extension $\LocRees{N,M_c'}$.
  \item Every leaf is labeled by a group which divides $M$.
\end{itemize}
In~\cite{Walter16diss}, it was shown that if $M$ is a monoid having $n$
elements which are not units, then there exists a decomposition tree of $M$
having at most $\Oh(3^{n/3})$ nodes.
However, it is not clear whether this bound optimal. Actually, it is not even known whether the size of the tree be bounded by a polynomial function.
Regardless of whether tight bounds can be obtained in the general case, it
would also be interesting to analyze subclasses of $\bfMon$. For example, it
is easy to see that for commutative monoids with a fixed number of generators,
there indeed is a polynomial bound.  What happens if the number of generators
is not fixed?

The starting point of our journey was the characterization of $\SD$ and $\SF$ by
aperiodic monoids. Having this theme in
mind, another interesting question  about the limits of the method arises. In~\cite{str79b}, Straubing showed
that the so-called Mal'cev product of $\AP$ and a group variety $\bfH$, denoted
by $\AP \malcev \bfH$, corresponds to the closure of $\bfH(A^*)$ under
concatenation product.
Following the proof of $\SD = \AP$ using local divisors, it is tempting to ask
whether the local divisor technique can also be applied to obtain a new,
possibly more general proof of Straubing's result.
In particular, it would be interesting to see whether there is a natural
language characterization of $\AP \malcev \bfH$ that relies on prefix codes
with bounded synchronization delay.
A major obstacle to initial attempts is a result of
Steinberg~\cite{Steinberg05} that $\AP \malcev \bfH$ is strictly contained in
$\ov \bfH$ for all non-trivial group varieties $\ov \bfH$.

\newcommand{\Ju}{Ju}\newcommand{\Ph}{Ph}\newcommand{\Th}{Th}\newcommand{\Ch}{Ch}\newcommand{\Yu}{Yu}\newcommand{\Zh}{Zh}\newcommand{\St}{St}\newcommand{\curlybraces}[1]{\{#1\}}

\end{document}